\documentclass[10pt, superscriptaddress,twocolumn, prl]{revtex4}

\usepackage{amsmath}
\usepackage{amssymb}
\usepackage{dsfont}
\usepackage{color}
\usepackage{hyperref}
\usepackage{graphicx}
\usepackage{epstopdf}
\usepackage{soul}
\usepackage{color}
\usepackage{braket}

\begin{document}
\title{Observation of localized flat-band modes in a quasi-one-dimensional \\
photonic rhombic lattice}


\author{Sebabrata Mukherjee}
\email{snm32@hw.ac.uk}
\affiliation{SUPA, Institute of Photonics and Quantum Sciences, Engineering $\&$ Physical Sciences, Heriot-Watt University, Edinburgh, EH14 4AS, United Kingdom}
\author{Robert R. Thomson}
\affiliation{SUPA, Institute of Photonics and Quantum Sciences, Engineering $\&$ Physical Sciences, Heriot-Watt University, Edinburgh, EH14 4AS, United Kingdom}



\begin{abstract}
We experimentally demonstrate the photonic realization of a dispersionless flat-band in a quasi-one-dimensional photonic lattice fabricated by ultrafast laser inscription. In the nearest neighbor tight binding approximation, the lattice supports two dispersive and a non-dispersive (flat) band. We experimentally excite superpositions of flat-band eigen modes at the input of the photonic lattice and show the diffractionless propagation of the input states due to their infinite effective mass. In the future, the use of photonic rhombic lattices, together with the successful implementation of a synthetic gauge field, will enable the observation of Aharonov–Bohm photonic caging.
\end{abstract}


\maketitle

{\it Introduction}
The dynamics of electrons in a crystal reveals many interesting phenomena that depend on the lattice geometry, external fields, presence of disorders and inter-particle interactions. The propagation of light waves across a photonic lattice, a periodic array of coupled optical waveguides, mimics the time evolution of the electronic wavefunction in a periodic potential. Due to this mapping, the photonic analogue of various solid-state-phenomena \cite{corrielli2013fractional, chiodo2006imaging, dreisow2009direct, dreisow2009bloch, rechtsman2013photonic} can be realized in the system of engineered waveguide-arrays. As with cold atoms in optical lattices \cite{bloch2005ultracold, jaksch2005cold, bloch2012quantum}, this artificial system allows us to engineer and access a desired Hamiltonian, and hence acts as a powerful platform for the study of various complex quantum mechanical effects in a clean environment. Indeed photonic lattices are ideal systems to study various effects in the absence of undesired excitations such as phonons in a real solid.

Localization effects of electronic wavefunctions in a crystalline solid have been studied for long time. In recent years, such effects have been experimentally observed in photonic systems. 
Localization effects due to the presence of disorder \cite{schwartz2007transport, martin2011anderson}, analogous external fields \cite {longhi2006observation, mukherjee2015modulation} and nonlinearity \cite {szameit2006two} have been observed. Recently, a novel type of localization due to a non-dispersive band in the energy spectrum was demonstrated in a photonic Lieb lattice \cite {mukherjee2015observation, vicencio2015Observation}. This localization effect which occurs due to the excitation of degenerate eigen modes can also be observed in quasi-one-dimensional lattices; see \cite{tasaki2008hubbard, Derzhko2012Low, johansson2015compactification}. In fact, one-dimensional photonic lattices are simpler to study theoretically, easier to fabricate experimentally and suitable for the study of many fundamental concepts. In this letter, we present a quasi-one-dimensional photonic rhombic lattice which supports two dispersive and a non-dispersive (flat) band.
We experimentally excite a superposition of flat-band eigen modes at the input of the lattice and show the diffractionless propagation of the input states due to their infinite effective mass. As also proposed by S. Longhi in Ref. \cite{longhi2014aharonov}, we suggest that such lattices, when implemented with a synthetic gauge field, will allow the observation of Aharonov–Bohm photonic caging.

\begin{figure}[h!]
\centering
{\includegraphics[width=\linewidth]{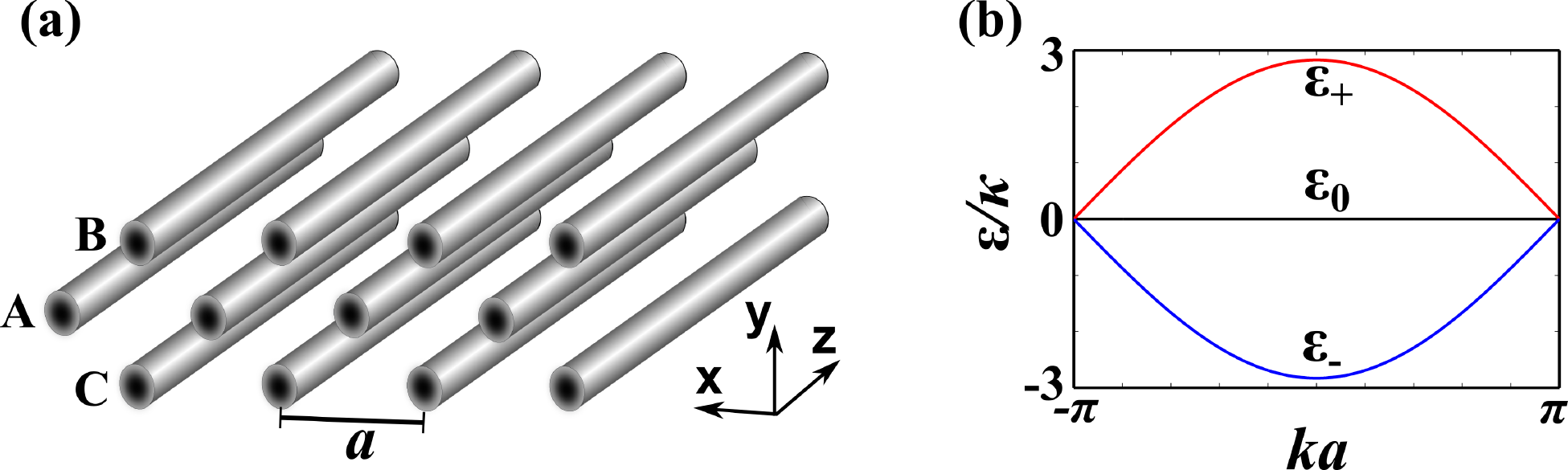}}
\caption{(a) A quasi-one-dimensional photonic rhombic lattice where the unit cell contains three lattice sites ($A, B$ and $C)$. The lattice constant is $a$. (b) In the nearest neighbor tight binding model, the lattice supports three energy bands. The upper and the lower bands are dispersive. The middle one is perfectly flat. The span of the Brillouin zone: ($-\pi \le ka \le \pi$). At $k=\pm \pi/a$ the three bands intersect each other.}
\label{fig1}
\end{figure}
\begin{figure}[t]
\centering
{\includegraphics[width=\linewidth]{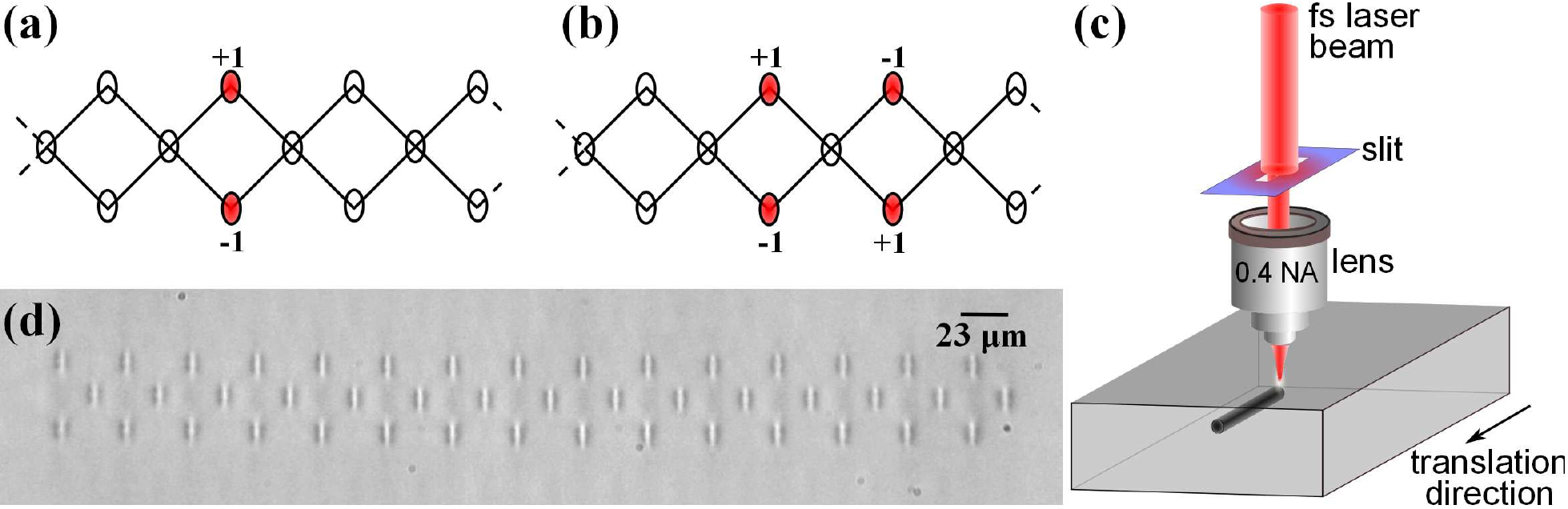}}
\caption{(a) A superposition of flat-band eigen modes can be excited at the input of the lattice if the $B$ and $C$-sites of a unit cell are excited with equal intensities ($I_B=I_C$) and opposite phases ($\phi_B=\phi_C\pm \pi$). See also Fig. \ref{fig3} (b and c). (b) The flat-band modes can also be excited if the $B$ and $C$-sites of two neighboring unit cells are excited with equal intensities ($I_B=I_C$) and alternating phases ($\phi_B=\phi_C\pm \pi$). See also Fig. \ref{fig3} (e and f). (c) Ultrafast laser inscription technique. The slit-beam shaping method was used to shape the refractive index profile, see \cite{davis1996writing, ams2005slit}. (d) White light transmission micrograph of the facet of a finite photonic rhombic lattice with fifteen unit cells. Each waveguide was observed to support a well confined fundamental mode at 780~nm wavelength with mode field diameters 8.3 $\mu$m and 7.0 $\mu$m along the $y$-axis (vertical) and $x$-axis (horizontal), respectively.}
\label{fig2}
\end{figure}
\begin{figure}[t]
\centering
{\includegraphics[width=\linewidth]{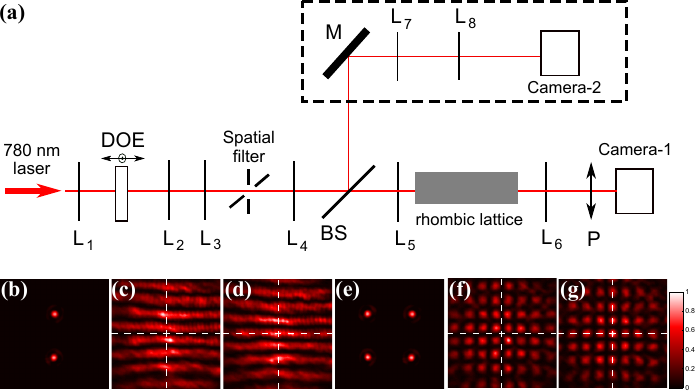}}
\caption{ (a) Experimental set-up for exciting the desired lattice sites at the input of the photonic rhombic lattice; see also \cite{mukherjee2015observation}. L$_1$-L$_8$ are convex lenses, M is a silver-coated mirror and BS is a beam splitter. A zero-order
nulled diffractive optical element (DOE) generates a square array of diffraction-order ``spots". A rectangular spatial filter (with adjustable height and width) blocks all the higher order diffraction ``spots" except for the first order ``spots". The spatial filter also helps to launch the desired number of first order spots. The separation among the four Gaussian spots was varied by translating the DOE along the propagation direction (z) of the laser beam.The relative phases of the four ``spots" can be controlled by translating the DOE in the x-y plane. The output intensity distribution is measured in Camera-1. A polarizer (P) passes only vertically polarized light. The set-up inside the dotted rectangle is used to observe the interference pattern of the input modes in the Fraunhofer regime.
(b and e) Two different input intensity distributions used in the experiment to excite flat-band eigen modes. The measured values of relative standard deviation (RSD) of the intensities are 3.9$\%$ and 4.1$\%$ respectively.
(c and f) The interference pattern of the flat-band modes in the Fraunhofer regime. (d and g) The interference pattern of the equal phase modes in the Fraunhofer regime.}
\label{fig3}
\end{figure}

{\it Photonic rhombic lattice}
In the paraxial approximation, the light propagation in a one dimensional photonic lattice is governed by the following Schr\"odinger-like equation \cite{longhi2009quantum, garanovich2012light}:
\begin{eqnarray}
i\lambdabar \frac {\partial \Psi(x, z)}{\partial z} & =&\Big[-\frac{\lambdabar^2}{2n_0} \frac {\partial ^2}{\partial x^2}-\Delta n(x) \Big] \Psi(x, z) \label{1}
\end{eqnarray}
where $\Psi (x, z)$ is the electric field envelop, $\lambda=2\pi\lambdabar$ is the free-space wavelength, $n_0$ is the average refractive index of the medium, the transverse refractive index profile [$\Delta n(x)$] acts as the effective potential for the light waves and $z$ plays the role of time. For well confined, single-mode waveguides one can use tight binding model to solve Eq. (\ref{1}).
Here we consider a quasi-one-dimensional photonic rhombic lattice as shown in Fig. \ref{fig1} (a). The unit cell contains three lattice sites ($A, B$ and $C)$. When only the lowest Bloch bands are excited Eq. (\ref{1}) gives the following coupled-mode equations:
\begin{eqnarray}
i \frac {\partial A_n}{\partial z} & =& \kappa \big(B_n+B_{n-1}+C_n+C_{n-1} \big) \nonumber \\
i \frac {\partial B_n}{\partial z} & =& \kappa \big(A_n+A_{n+1} \big) \nonumber \\
i\frac {\partial C_n}{\partial z} & =& \kappa \big(A_n+A_{n+1} \big) \label{2}
\end{eqnarray}
where $A_n$, $B_n$ and $C_n$ are the electric field amplitudes at the $A$, $B$ and $C$ sites of the $n$-th unit cell respectively, $\kappa$ is the nearest-neighbor  hopping amplitude (or coupling constant). The intensity distribution at the output of the photonic lattice, for a given input excitation, is obtained by solving Eq. (\ref{2}).
By diagonalizing the Fourier transformed Hamiltonian, the following dispersion relations are obtained \cite{vidal2000interaction, longhi2014aharonov}:
\begin{eqnarray}
\varepsilon_\pm (k)&=&\pm2\kappa\sqrt{1+\cos(ka)} \nonumber \\
\varepsilon_0 (k) &=&0  
\label{3}
\end{eqnarray}
where $a$ is the lattice constant, $\varepsilon_0$ is the energy of non-dispersive (flat) band, $\varepsilon_\pm$ represent the upper and the lower dispersive bands respectively. The span of the Brillouin zone is given by ($-\pi \le ka \le \pi$). At $k=\pm \pi/a$ the three bands intersect each other; see Fig. \ref{fig1} (b). 
A superposition of flat-band eigen modes can be excited at the input of the lattice if (a) the next-nearest neighbor coupling is insignificant and (b) the $B$ and $C$ sites of a unit cell are excited with equal intensities ($I_B=I_C$) and opposite phases ($\phi_B=\phi_C\pm \pi$); see Fig. \ref{fig2} (a). 

\begin{figure}[t]
\centering
{\includegraphics[width=\linewidth]{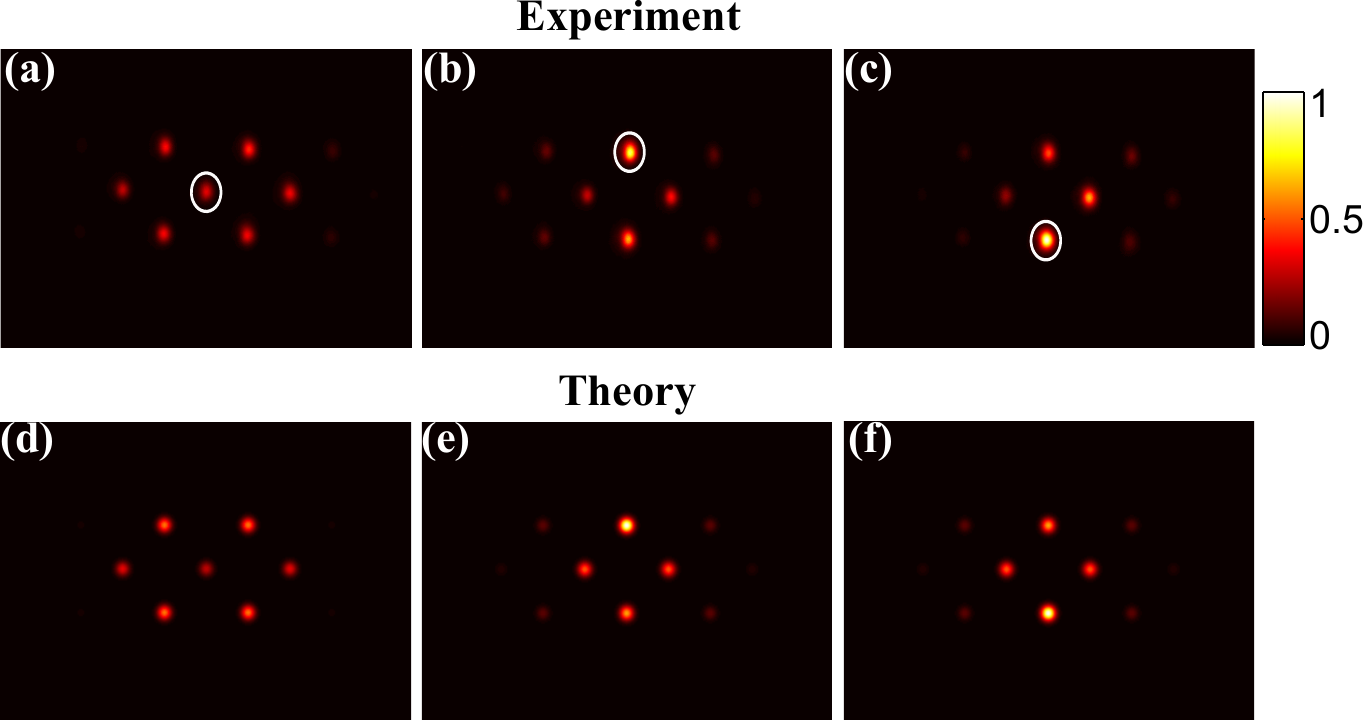}}
\caption{(a-c) Experimentally observed intensity distributions at the output of a 70-mm-long photonic rhombic lattice when $A, B$ and $C$-sites (indicated by the white circles) were excited at the input respectively. The field of view is approximately 110 $\mu$m by 150 $\mu$m. The simulated output intensity distributions for $A, B$ and $C$-site excitation are shown in (d-f) respectively. The images are normalized such that the total power for each image is 1.}
\label{fig4}
\end{figure}

Twelve sets of photonic rhombic lattices (waveguide-to-waveguide separation, $a/\sqrt{2}$ = 15 $\mu$m to 26 $\mu$m in steps of 1 $\mu$m) were fabricated inside a 70-mm-long glass (Corning Eagle$^{2000}$) substrate using ultrafast laser inscription \cite{davis1996writing, ams2005slit}, see Fig. \ref{fig2} (c). Each waveguide was fabricated by translating the glass substrate once through the focus of a femtosecond laser beam at 8~mm-s$^{-1}$ translation speed. The fabrication parameters were optimized to inscribe well-confined single-mode waveguides at 780~nm wavelength; see \cite{mukherjee2015observation} for detailed fabrication parameters.
Each lattice contains fifteen unit cells. A white light transmission micrograph of the facet of a lattice is shown in Fig. \ref{fig2} (d). All the experiments were performed using the lattice with $a/\sqrt{2}=23$~$\mu$m, for which $\kappa=0.01$~mm$^{-1}$ and the next nearest-neighbor coupling constant ($\kappa_n$) was insignificant over a length of 7 cm. 
The ratio of nearest neighbor and next-nearest neighbor coupling strength ($\kappa/\kappa_n$) was estimated to be $\approx 12$. The value of $\kappa_n$ was obtained by measuring the coupling strength of twelve sets of evanescently coupled two-waveguide couplers (inter-waveguide separations: 11~$\mu$m to 22~$\mu$m in steps of 1~$\mu$m) and optimally fitting the variation to an exponentially decaying function; see \cite{corrielli2013fractional, szameit2007control}.

{\it Optical excitation}
The experimental setup to excite the desired lattice sites at the input of the photonic rhombic lattice is shown in Fig. \ref{fig3} (a). First order diffraction ``spots" generated by a zero-order nulled diffractive optical element (DOE) were used to launch four Gaussian spots with almost equal intensities (the measured relative standard deviation (RSD)$\approx 4\%$). The spatial filter blocks all higher order diffraction ``spots". It also helps to pass any desired number of first order spots. The separation among the four Gaussian spots was varied by translating the DOE along the propagation direction (z) of the laser beam.The relative phases of the four ``spots" can be controlled by translating the DOE in the x-y plane. The relative phases among the Gaussian ``spots" were measured by observing the interference pattern in the Fraunhofer regime; see the set-up inside the dotted rectangle in Fig. \ref{fig3} (a). The polarizer (P) in front of Camera-1 passes only the vertically polarized (which is one of the eigen polarization of the waveguides) light. Hence our measurements are not affected by the polarization dependent coupling.

\begin{figure}[t]
\centering
{\includegraphics[width=\linewidth]{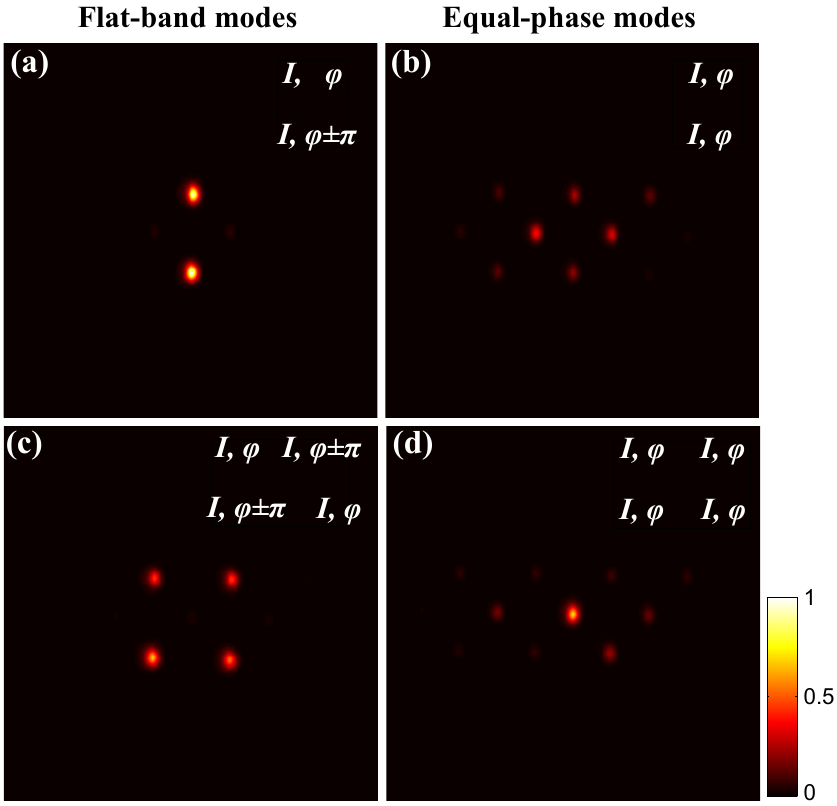}}
\caption{Experimentally observed flat-band modes (a, c) and the equal-phase modes (b, d). The intensities and relative phases of the input modes are shown on the top-right of each image. The output intensity distribution when the $B$ and $C$-sites of a unit cell were excited with equal intensities, opposite (a) and equal (b) phases. When the flat-band mode is excited by coupling light into the $B$ and $C$-sites, light can not tunnel to the nearest $A$-sites due to destructive interference. Localized flat-band modes can also be observed by exciting the $B$ and $C$-sites of two neighboring unit cells with equal intensities and alternating phases (c). When these four $B$ and $C$-sites were excited with equal intensities and equal phase, the mode in not localized (d). The field of view is approximately 160 $\mu$m by 160 $\mu$m. The images are normalized such that the total power for each image is 1.}
\label{fig5}
\end{figure}

Fig. \ref{fig4} (a-c) show the output intensity distributions when $A$, $B$ and $C$-sites were excited individually at the input respectively. Fig. \ref{fig4} (d-f) show the corresponding simulated  intensity distributions obtained by solving Eq. (\ref{2}). The fluctuation in the nearest neighbor coupling constant (due to random variations in the waveguide-to-waveguide separation, in other words off-diagonal disorder) was estimated by comparing the observed and simulated output intensity distributions and found to be $\approx 0.001$ mm$^{-1}$. It should be noted that the light contained at the initially excited $A$-site (indicated by the white circle) is less than that at $B$ (and/or $C$)-site. In other words, light coupled to $A$-sites diffracts more than light coupled to $B$ (and/or $C$)-sites. Indeed the calculation of the overlap between the input state and the eigen states of the three bands show that the  $A$-site excitation excites more eigen states from the dispersive bands compared to $B$ (and/or $C$)-site excitation.

To excite the flat-band modes, we couple light into the $B$ and $C$-sites of a unit cell with equal intensity and opposite phases (the flat-band mode). We observe that the light remains localized to the excited sites without tunneling to other waveguides; see Fig. \ref{fig5} (a). This localization effect is due to the interference effect. When the flat-band modes are excited by coupling light into the $B$ and $C$-sites, light can not tunnel to the nearest $A$-sites due to destructive interference. It should be noted that this can only occur if the next-nearest neighbor coupling is insignificant. 
To ensure the sensitivity of phase profile, we couple light into the $B$ and $C$-sites with equal intensities and equal phases (the equal-phase mode). As shown in Fig. \ref{fig5} (b), this input mode is not localized.

The above mentioned input mode is not unique to excite the flat-band modes. In the nearest neighbor approximation, the flat-band modes can also be excited if the B and C-sites of two neighboring unit cells are excited with equal intensities ($I_B=I_C$) and alternating phases ($\phi_B=\phi_C\pm \pi$); see Fig. (\ref{1}f). The localized output intensity distribution for this second flat-band mode is shown in Fig. \ref{fig5} (c). 
In the next step we excited these four waveguides with equal intensities and equal phase (the second equal-phase mode). These modes are not localized [Fig. \ref{fig5} (d)] as would be expected.

\begin{figure}[t]
\centering
{\includegraphics[width=\linewidth]{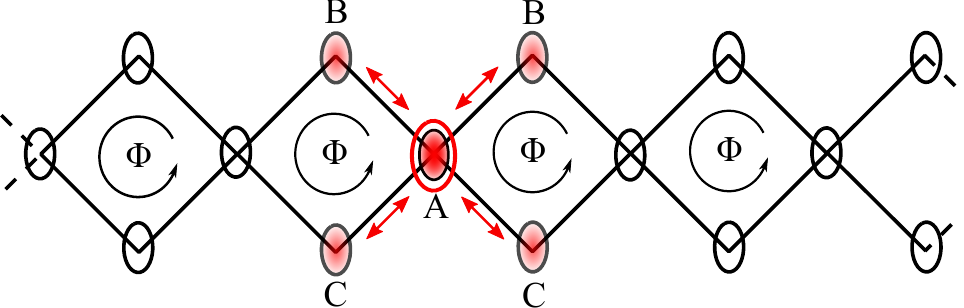}}
\caption{Aharonov-Bohm caging in a photonic rhombic lattice in the presence of a synthetic magnetic flux, $\Phi=\pi$ per plaquette; see Ref. \cite{longhi2014aharonov}. When a single $A$-site is excited at the input, the light intensity at the excited lattice site will periodically oscillate between 1 and 0 along the propagation ($z$) direction.  Due to photonic Aharonov-Bohm caging, the light intensity is locked into the initially excited $A$-site and its nearest neighbor sites (four $B$ and $C$-sites).}
\label{fig6}
\end{figure}

{\it Aharonov-Bohm photonic caging}
In comparison to two dimensional photonic lattices, the rhombic lattice we have investigated here is a greatly simplified system which can be used to study complex phenomena. In particular, synthetic gauge fields are one of the exciting routes to engineer the band structure of photonic lattices. In the presence of magnetic fields, various interesting phenomena, such as, the Hofstadter spectrum and the fractal quantum Hall effect \cite{dean2013hofstadter} and topologically protected chiral edge modes \cite{Kitagawa2010Topological, fang2012realizing, rechtsman2013strain, longhi2013effective}, can be observed in suitable lattice geometries. A photonic rhombic lattice can be used in conjunction with synthetic gauge fields, to observe Aharonov-Bohm photonic caging, a magnetic field induced localization effect which occurs due to the destructive interference at some specific values of magnetic flux; see \cite{vidal2000interaction, longhi2014aharonov, Abilio2014magnetic}. 
The experimental implementation of a synthetic gauge field in a photonic lattice is an experimental challenge, and it requires an engineered lattice where a photon tunneling along a closed loop on the lattice will acquire a non-vanishing phase which is analogous to an Aharonov-Bohm phase.
In principle it can be realized by generating complex-valued hopping amplitudes. As proposed in Ref. \cite{longhi2014aharonov}, a circular bending and periodic modulation of propagation constant (which can be realized by changing the translation speed) of the waveguides along the $z$ direction would generate a magnetic flux in a photonic rhombic lattice. Considering a rhombic lattice [see Fig. \ref{fig6}] in a gauge field with $\Phi$ flux per plaquette, the dispersion relation becomes \cite{vidal2000interaction, longhi2014aharonov}: 

\begin{eqnarray}
\varepsilon_\pm (k)&=&\pm2\kappa\sqrt{1+\cos(\Phi/2) \cos(ka-\Phi/2)} \nonumber \\
\varepsilon_0 (k) &=&0  
\label{4}
\end{eqnarray}

For $\Phi=\pi$, the energy spectrum consists of three non-dispersive (flat) bands. In this situation, if a single $A$-site is excited at the input, the light intensity at the excited lattice site will periodically oscillate between 1 and 0 along the $z$ direction.  Due to photonic Aharonov-Bohm caging, the light intensity is locked into the initially excited $A$-site and its nearest neighbor sites (four $B$ and $C$-sites).

{\it Conclusion}
In conclusion, we have presented a quasi-one-dimensional photonic rhombic lattice and experimentally excite a superposition of eigenmodes of the non-dispersive band of the lattice. We show that the flat-band modes remains localized to the initially excited waveguides. The reason for this effect is that the eigenmodes in the flat band are degenerate, so any superposition of them behave as an eigen mode. This type of localization effect is sensitive to the modes excited at the the input which makes this effect different from the disorder or nonlinearity induced localizations. The results presented in this letter can be regarded as a first step of lattice band-engineering by a synthetic gauge field.
Flat bands may be useful for diffractionless image transport \cite{vicencio2014diffraction}.
In the presence of nonlinearities we also expect novel types of nonlinear
dynamics due to the absence of any kinetic terms in the effective Hamiltonian.

\bigskip
\noindent
\textbf{Acknowledgments.} We are pleased to thank E. Andersson, N. Goldman, P. \"Ohberg, A. Spracklen and M. Valiente for helpful discussions. We also thank Alexander Arriola, Debaditya Choudhury and David G. MacLachlan for building the initial fabrication set-up.

\bigskip
\noindent
\textbf{Funding.} UK Science and Technology Facilities Council (STFC) in the form of an STFC Advanced Fellowship (ST/H005595/1). James Watt Scholarship from Heriot-Watt University.



\end{document}